\documentstyle[epsfig]{aipproc}
\begin{document}

\rightline{Report No. HD-THEP-00-56}

\title{ELECTROGENESIS IN A SCALAR FIELD DOMINATED EPOCH}

\author{Tomislav Prokopec}
\address{Universit\"at Heidelberg, Institut f\"ur Theoretische Physik,\\
Philosophenweg 16, D-69120 Heidelberg, Deutschland}

\maketitle

 \begin{abstract}
 In this talk I discuss models in which a homogeneous scalar field
is used to modify standard cosmology above the nucleosynthesis scale 
to provide an explanation for the observed
matter-antimatter asymmetry of the Universe.
 \end{abstract}

\section{Introduction}

 Scalar fields are used to model either a very early universe ({\it inflation}),
or a very late universe ({\it `quintessence'})~\cite{jf-zws}. Their role at
intermediate times is however largely ignored. In this 
talk~\footnote{Invited talks presented at the conferences 
{\it Strong and Electroweak Matter} (SEWM-2000) in Marseille, (June 14-17, 2000),
and {\it Cosmology and Astroparticle Physics (CAPP-2000)} in Verbier, Switzerland 
(July 17-28, 2000); based on work with Michael Joyce
\cite{JoyceProkopec,Prokopec,JoyceProkopec2}.} I discuss models in which 
a homogeneous scalar field is used to modify the standard cosmology at the 
electroweak scale to allow for baryogenesis, even when the electroweak
transition is smooth or weakly first order. 

 In order to produce any baryon number, a {\it source} is required that
drives the Universe out of equilibrium~\cite{ShaposhnikovRubakov}. 
Since at the electroweak scale the 
expansion rate is very small ($H/T\sim 10^{-16}$), it is often assumed that 
it cannot drive baryogenesis, simply because the produced baryon-to-entropy ratio, 
$n_B/s\propto H/T$, is too small to account for observation, 
$(n_B/s)_{\rm observed}\sim 5\times 10^{-11}$. 

\section{Scalar fields and the expansion rate}

\begin{figure}[h!]
\vspace*{-3mm}
\centerline{\epsfxsize = 3.5in  \epsffile{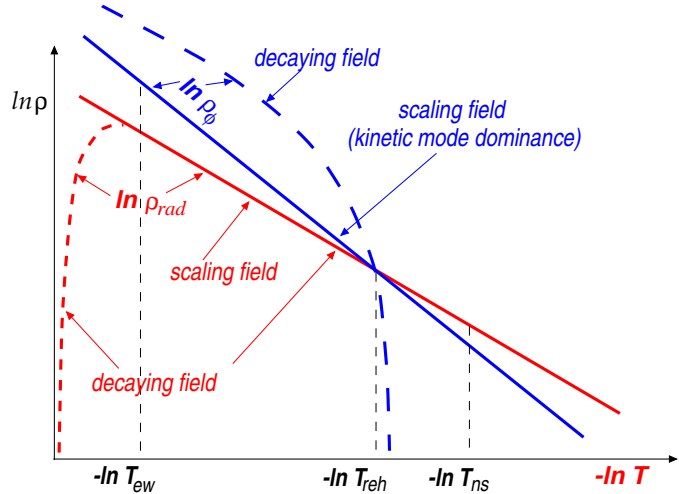}}
\vspace*{10pt}
\caption[fig1]{\label{fig1} Evolution of energy density in radiation and 
the dominant scalar field as a function of temperature. 
Two cases are illustrated: 
{\it Model (A)} in which the dominant scalar component scales faster
then radiation, but does not decay ({\it solid lines}), and
{\it Model (B)} in which the scalar field decays ({\it dashed lines}).}
\end{figure}

In Refs.~\cite{JoyceProkopec}
we discussed models in which, based on the dominance of a kinetic scalar
field mode ({\it kination}), the expansion rate of the Universe changes to
\begin{equation}
\frac{H}{T} \sim \frac{T}{T_{\rm reh}}\left(\frac{H}{T}\right)_{\rm rad}
\,,
\label{H}
\end{equation}
where $({H}/{T})_{\rm rad}$ is the expansion rate in radiation-dominated 
universe, and $T_{\rm reh}$ is the `reheat' temperature at which 
the energy-densities are equal,
$\rho_{\rm \phi}\sim \rho_{\rm rad}$ (see {\it Model A} in figure~\ref{fig1}). 
At the nucleosynthesis scale, $T_{\rm ns}\sim 1$MeV, the Universe is radiation
dominated, and hence $T_{\rm reh}\geq T_{\rm ns}$.
Since $T_{\rm ew}/T_{\rm ns}\sim 10^5$, the expansion rate at the electroweak scale
($T_{\rm ew}\sim 100$GeV) can be enhanced to about
$({H}/{T})_{\rm ew}\sim 10^{-11}$. With some tuning in the parameters of 
the model~\cite{JoyceProkopec}, this is enough for successful baryogenesis 
even at a smooth or a weakly first order transition. We also note 
that the same scalar field can be both the inflaton and the 
kinaton~\cite{JoyceProkopec}.

 The expansion rate can be further enhanced if the scalar field decays. In this case
we have 
\begin{equation}
\frac{H}{T} \sim \left(\frac{T}{T_{\rm decay}}\right)^2
              \left(\frac{H}{T}\right)_{\rm rad}
\,,
\label{H2}
\end{equation}
where again $T_{\rm ew}\geq T_{\rm decay}\geq T_{\rm ns}$, and 
$T_{\rm decay}=T_{\rm reh}$ is the temperature at which the field decays 
(see {\it Model B} in figure~\ref{fig1}). At a first sight one can get the 
expansion rate at the
electroweak scale high enough to drive baryogenesis. There is a caveat however:
as the scalar field decays, entropy is released, which then dilutes the original
baryon number produced at the electroweak scale. The entropy release is minimal
if the scalar field energy is dominated by the kinetic mode~\cite{Prokopec}.

\section{Approximately conserved charges and baryon number} 

 We have argued that the dominance of a kinetic scalar mode can be used to increase
the expansion rate of the Universe at the electroweak scale by orders of
magnitude. In fact the expansion rate can easily become larger than the 
destruction rate of some of approximately conserved species, {\it e.g.}
\begin{equation}
H(T_{\rm ew})\geq \Gamma_{e_R},\Gamma_{\mu_R},\Gamma_{u_R}, ..
\label{H-Gamma}
\end{equation}
This simply means that, if any of these charges are produced above the electroweak
scale, they decay only below the electroweak scale, that is when the baryon
number violating processes are already frozen-in. 
If a net right-handed electron number, $e_R$, is produced at a scale
$T>T_{\rm ew}$, in chemical equilibrium the baryon number $B$ is shifted
to~\cite{JoyceProkopec2}
\begin{equation}
B\approx \frac{1}{3}e_R
\,.
\label{B-eR}
\end{equation}
This simple relation describes the correct local chemical equilibrium 
as long as the destruction rate for the right-handed electrons is large when
compared with the expansion rate at the electroweak scale, {\it i.e.}
$\Gamma_{e_R}\sim 10^{-13}T_{\rm ew}<H(T_{\rm ew})$. 
At $T=T_{\rm ew}$ the baryon-number violating processes (`sphalerons') 
freeze-in, {\it i.e.} the sphaleron rate drops below the expansion rate,
and the baryon number~(\ref{B-eR}) remains frozen until today. Eq.\ (\ref{B-eR}) 
can be intuitively understood as follows. Above the electroweak scale 
the Universe must be hypercharge-neutral, $Y=0$. In the presence of net $e_R$ the
corresponding electron hypercharge $Y_{e_R}=y_{e_R} e_R$ must be screened by 
various charges pulled out of the plasma. The charges that minimize 
the relevant free energy include the quarks that carry a net
baryon number $B$ as given in Eq.\ (\ref{B-eR}).

\section{Electrogenesis} 

 We now present a simple perturbative model for production of the right-handed
electrons required for baryogenesis ({\it cf.} Eq.~(\ref{B-eR})). 
To this purpose we introduce heavy scalar fields
$\Phi^a$ (with a mass at least in the TeV range) that couple to the standard-model 
fermions {\it via} a Yukawa interaction term of the form 
\begin{equation}
{\cal L}_{CP}[\Phi^a]  = - h^a_{ij}\Phi^a \bar\psi_{iL}\psi_{jR} + h.c.\,,
\end{equation}
where the couplings  $h^a_{ij}$ are CP violating, {\it i.e.} 
${\bf h^a}^\dag\neq {\bf h^a}$ (${\bf h^a}$ denotes the matrix of couplings).
In order to violate CP symmetry, ${\bf h^a}$ must contain 
a complex phase unremovable by phase transformations on the whole Lagrangian, 
which can be achieved by the flavor mixing structure and
the existence of at least two such scalars. 
The most stringent constraints on the masses and the couplings of such scalars 
come from the fact that they are flavor changing. For leptons
the strongest constraint of this type comes from the bounds on the
decay $\mu \rightarrow e \gamma$. For couplings $h^a_{ij}$ of
order one this requires masses $M_\Phi \geq 100$TeV.

\begin{figure}[h!] 
\vspace*{-3mm}
\centerline{\epsfxsize = 4in  \epsffile{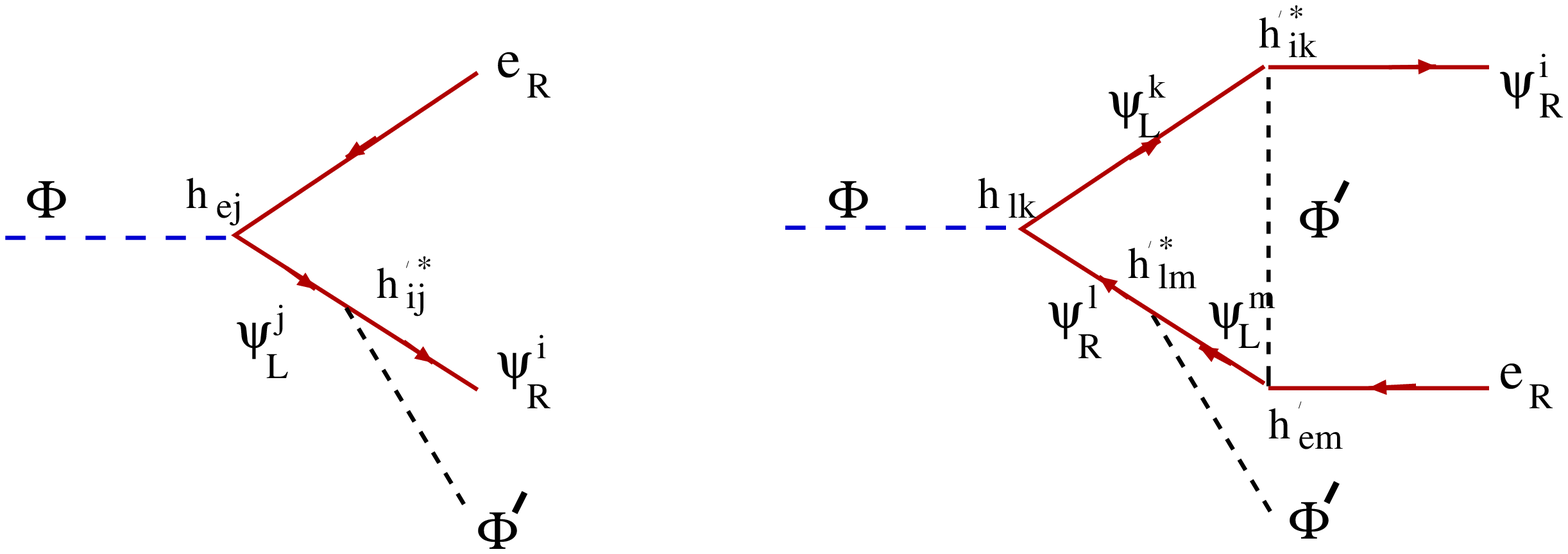}}
\vspace*{10pt}
\caption[fig2]{\label{fig2} Tree and one loop diagrams for the
three body decay $\Phi \rightarrow \bar{e}_R \Psi^i_R \Phi'$,
with the appropriate couplings at the vertices. 
We assume that $\Phi$ is heavier than $\Phi'$. 
When the second outgoing lepton is a $\mu$ or $\tau$ lepton
the process produces net $e_R$ number.}
\end{figure}
 When $\Phi^a$ decay out of equilibrium, a net $e_R$ may be produced.
An example of such a decay channel is shown in figure~\ref{fig2}, where 
CP violation is realised as the interference term between the tree level 
and 1-loop 3-body decay channels ({\it cf.} \cite{YanagidaYoshimuraFukugita}). 
The resulting electron-to-entropy ration is then 
\begin{equation}
\frac{e_R}{s} \sim \frac{10^{-2}}{g_*}\vert {\bf h}\vert^4 \delta_{\rm CP}\,,
\label{eR asymmetry}
\end{equation}
where $g_*$ is the number of relativistic degrees of freedom in the plasma, 
$\delta_{\rm CP}$ the relevant CP-violating angle. 
In order that $\Phi^a$ decay out of equilibrium, they ought to be sufficiently 
massive. One finds~\cite{JoyceProkopec2} that in {\it Model A,}
\begin{equation}
M_\Phi
%> 3 \times 10^6 {\rm GeV} \vert{\bf  h}\vert
%      \left(\frac{T_{\rm reh}}{T_{\rm ns}}\right)^{\frac{1}{2}}
\,> \, 5 \, \vert{\bf  h}\vert \times 10^6 \,{\rm GeV}, 
\label{mass Phi A}
\end{equation}
while in {\it Model B,} when the dominant component decays, the bound reads 
\begin{equation}
 M_\Phi
%> 2\, \vert{\bf  h}\vert^{\frac{2}{3}}
%\,\left(\frac{T_{\rm reh}}{T_{\rm ns}}\right)^{\frac{2}{3}} \, {\rm TeV}
\,> \, 3 \, \vert{\bf  h}\vert^{\frac{2}{3}} \, {\rm TeV}.
\label{mass Phi B}
\end{equation}
This implies that the scalars $\Phi^a$ may be observable 
by the future accelerators ({\it e.g.} LHC). This fact alone
gives a sufficient motivation for a more detailed investigation of the models
that contain such heavy scalar fields.

%
%
%%%%%%%%%%%%%%%%%%%%%%%%%%%%%%%%%%%%%%%%%%%%%%%%%%%%%%%%%%%%%%%%%%%%%%%%%%%%%%%
%  References
%%%%%%%%%%%%%%%%%%%%%%%%%%%%%%%%%%%%%%%%%%%%%%%%%%%%%%%%%%%%%%%%%%%%%%%%%%%%%%%
%
%\section*{References}

\end{document}